\documentclass{aa}

\usepackage{natbib}

\bibpunct{(}{)}{;}{a}{}{,} % to follow the A&A style

\usepackage{graphicx}

\usepackage{amssymb,amsmath}
\usepackage{delarray}
\usepackage{mathrsfs}
\usepackage{multirow}
\usepackage{color}

\usepackage{balance}

\usepackage[pdftitle={Non-magnetic photospheric bright points in 3D simulations of the solar atmosphere},%
            pdfauthor={F. Calvo, O. Steiner and B. Freytag},%
            pdfsubject={Solar physics},%
            pdfkeywords={Sun: photosphere, Sun: granulation, Hydrodynamics, Turbulence},%
            unicode=true,%
            bookmarks=true,%
            bookmarksnumbered=false,%
            bookmarksopen=false,%
            breaklinks=false,%
            pdfborder={0 0 .5},%
            backref=false,%
            colorlinks=true,%
            citecolor=blue,%
            linkcolor=blue,%
            urlcolor=blue]{hyperref}
\usepackage[all]{hypcap}

\usepackage[varg]{txfonts}

\usepackage{etoolbox}
\makeatletter

\patchcmd{\NAT@citex}
{\@citea\NAT@hyper@{%
		\NAT@nmfmt{\NAT@nm}%
		\hyper@natlinkbreak{\NAT@aysep\NAT@spacechar}{\@citeb\@extra@b@citeb}%
		\NAT@date}}
{\@citea\NAT@nmfmt{\NAT@nm}%
	\NAT@aysep\NAT@spacechar\NAT@hyper@{\NAT@date}}{}{}

\patchcmd{\NAT@citex}
{\@citea\NAT@hyper@{%
		\NAT@nmfmt{\NAT@nm}%
		\hyper@natlinkbreak{\NAT@spacechar\NAT@@open\if*#1*\else#1\NAT@spacechar\fi}%
		{\@citeb\@extra@b@citeb}%
		\NAT@date}}
{\@citea\NAT@nmfmt{\NAT@nm}%
	\NAT@spacechar\NAT@@open\if*#1*\else#1\NAT@spacechar\fi\NAT@hyper@{\NAT@date}}
{}{}

\makeatother

\newcommand{\vect}[1]{\boldsymbol{#1}}
\newcommand{\N}{\mathcal{N}}
\newcommand{\I}{\mathcal{I}}
\newcommand{\Ic}{C_{I}}
\newcommand{\Rhoc}{C_{\rho}}
\newcommand{\Pc}{C_{P}}
\newcommand{\Tc}{C_{T}}
\newcommand{\VvMax}{v_\mathrm{v}^{\mathrm{max}}}
\newcommand{\VhMax}{v_\mathrm{h}^{\mathrm{max}}}
\newcommand{\Vort}{\omega_{v,z}}
\newcommand{\D}{D}
\newcommand{\WD}{W_\mathrm{d}}
\newcommand{\cobold}{CO\textsuperscript{5}BOLD}

\begin{document}
	
\AANum{AA/2016/28649}
	
\title{Non-magnetic photospheric bright points in 3D simulations of 
       the solar atmosphere\thanks{The movie associated to Fig.\,1 is 
       available at \newline \protect\url{http://www.arxiv.org}}}

\author{F.~Calvo\inst{1,2}
        \and O.~Steiner\inst{1,3}
        \and B.~Freytag\inst{4}}

\institute{Istituto Ricerche Solari Locarno (IRSOL), 
           via Patocchi 57\,--\,Prato Pernice, 6605 Locarno-Monti, Switzerland
           \and 
           Geneva Observatory, University of Geneva, 
           Ch. des Maillettes 51, 1290 Sauverny, Switzerland\\
           \email{\href{mailto:flavio.calvo@irsol.ch}{\color{magenta}{flavio.calvo@irsol.ch}}}
           \and Kiepenheuer-Institut f\"ur Sonnenphysik, 
           Sch\"oneckstrasse 6, 79104 Freiburg, Germany
           \and {Division of Astronomy and Space Physics, 
           Department of Physics and Astronomy, Uppsala University, 
           Box 516, SE-751 20 Uppsala, Sweden}
           }

\date{Received 6 April 2016 / Accepted 30 August 2016}

\abstract
{Small-scale bright features in the photosphere of the Sun, such as faculae
or \emph{G-}band bright points, appear in connection with small-scale magnetic flux
concentrations.
}
{Here we report on a new class of photospheric bright points that are free 
of magnetic fields. So far, these are visible in numerical simulations only.
We explore conditions required for their observational detection.
}
{Numerical radiation (magneto-)hydrodynamic simulations of the near-surface 
layers of the Sun were carried out. 
The magnetic field-free simulations show tiny bright points, reminiscent 
of magnetic bright points, only smaller. A simple toy model for these 
non-magnetic bright points (nMBPs) was established that serves as a base 
for the development of an algorithm for their automatic detection.
Basic physical properties of 357 detected nMBPs were extracted and 
statistically evaluated. We produced synthetic intensity maps that mimic 
observations with various solar telescopes to obtain hints on their 
detectability.
}
{The nMBPs of the simulations show a mean bolometric intensity contrast
with respect to their intergranular surroundings of approximately 20\%, a size 
of 60--80\,km, and the isosurface of optical depth unity is at their location
depressed by 80--100\,km. They are caused by swirling downdrafts that 
provide, by means of the centripetal force, the necessary pressure gradient 
for the formation of a funnel of reduced mass density that reaches from the 
subsurface layers into the photosphere. Similar, frequently occurring funnels 
that do not reach into the photosphere, do not produce bright points.
}
{Non-magnetic bright points are the observable manifestation of vertically
extending vortices (vortex tubes) in the photosphere. The resolving power of
4-m-class telescopes, such as the DKIST, is needed for an unambiguous detection 
of them.
}


\maketitle

%
%%%%%%%%%%%%%%%%%%%%%%%%%%%%%%%%%%%%%%%%%%%%%%%%%%%%%%%%%%%%%%%%%%%%%%%%%%%%%
%
\section{Introduction}
\label{sect:intro}
Brilliant minuscule features, visible in white light on the solar disk,
have attracted the attention of observers from the beginning of solar physics.
In the early days of solar observation, it was the faculae that were of interest, but with the
advent of high resolution observation, the ``solar filigree'' 
\citep{dunn+zirker1973} or ``facular points'' \citep{mehltretter1974}
have caught the attention of observers. Later, these objects became 
well-known as the ``\emph{G-}band bright points'' \citep{muller+roudier1984},
which were investigated in great detail with the Swedish Solar 
Telescope (SST), for example by
\citet{berger+title1996,berger+title2001} and \citet{berger+al2004}.
Both faculae and \emph{G-}band bright points host concentrations
of magnetic fields with a strength of approximately 1000 Gauss, which are indeed 
at the origin of the brightenings.
\citet{utz+al2014} give properties of individual- and the statistics of
an ensemble of 200 magnetic bright points, observed with the 1\,m balloon-borne solar telescope Sunrise.

Magnetohydrodynamic simulations have been carried out to explain
the brilliance, shape, and physics of  faculae 
\citep[e.g.][]{carlsson+al2004,keller+al2004,steiner2005,beeck+al2013,beeck+al2015}, 
the origin of \emph{G-}band bright points and their congruency with magnetic flux 
concentrations \citep[e.g.][]{shelyag+al2004,nordlund+stein+asplund2009} 
or properties of the solar filigree
\citep[e.g.][]{voegler+al2005,stein2012,moll+al2012}.
Various explanations on the extra contrast of \emph{G-}band bright points were 
provided by 
\citet{steiner+al2001}, \citet{sanchez_almeida+al2001}, or 
\citet{schuessler+al2003}.

Non-magnetic photospheric bright points have also been reported to exist
\citep{berger+title2001,langhans+al2002}. These features are likely
parts of regular granules, or 
the late stage of collapsing granules, 
or tiny granules. Turning to numerical simulations, \citet{moll+al2011} 
find in their simulations the formation of strong, vertically oriented 
vortices and they discuss a single event that gives rise to a depression 
in the optical depth surface $\tau = 1$ and a locally increased radiative 
bolometric intensity of up to 24\%, confined to within an area of less than 
100\,km in diameter. Since this event occurs in a simulation that contains only 
a weak magnetic field, their bright point is essentially non-magnetic in nature.
\citet[][Fig.\,6]{Freytag2013} shows a bolometric intensity map of the 
hydrodynamic (non-magnetic) high-resolution \cobold\ run 
\verb#d3t57g44b0#, which shows a number of brilliant tiny dots. This
run contributes part of the data that is  analyzed herein. 
The present paper focuses on the derivation of statistical properties 
of this type of bright point rather than on single events or their mechanism.

In this paper, we report on two magnetic field-free radiation-hydrodynamics 
simulations, which show tiny bright features of about 50--100\,km in size,
located in the intergranular space, well separated from granules, yet
non-magnetic by construction. They obviously represent another class of 
photospheric bright feature, which we henceforth refer to as 
\emph{\textit{non-magnetic bright points}} (nMBPs).

From these simulation data, we extract a total of 357 nMBPs from which we derive 
statistical properties and we investigate conditions for the observational 
detection of them. In Sect.~\ref{sect:nMBP}, a representative nMBP is displayed and we propose a 
theoretical toy model for it that provides us the basis for a recognition 
algorithm that is given in Sect.~\ref{sect:toy}. The basic physical properties and statistics
of the nMBPs are presented in Sect.~\ref{sect:pysprop}. In
Sect.~\ref{sect:synthobs} we look at synthetic intensity maps and
discuss conditions for the observation of nMBPs. Conclusions are
given in Sect.~\ref{sect:conclusion}.

%
%%%%%%%%%%%%%%%%%%%%%%%%%%%%%%%%%%%%%%%%%%%%%%%%%%%%%%%%%%%%%%%%%%%%%%%%%%%%%
%
\section{nMBPs from the simulations}
\label{sect:nMBP}
The magnetic field-free, three-dimensional simulations that we have carried 
out cover a horizontal section (field-of-view) of $9600\times 9600\,\mbox{km}^2$ 
of the solar atmosphere, extending from the top of the convection zone 
beyond the top of the photosphere over a height range of $2800\,\mbox{km}$.
The optical depth surface $\tau_{500\mbox{\scriptsize nm}}=1$ is located approximately in the 
middle of this height range. The spatial resolution of the computational
grid is $10\times 10\times 10\,\mbox{km}^{3}$ (size of a computational cell), 
identical in all directions. A full data cube of the simulation (full box)
is stored every 4~min, whereas a two-dimensional map of the vertically
directed, bolometric radiative intensity at the top boundary is produced every
10~s.

The simulations were carried out with the \cobold\ code, 
a radiative magneto-hydrodynamics code described in \cite{Freytag2012}.
For our present study of nMBPs, we use the run carried out with the 
hydrodynamic Roe module, which implements an approximate Riemann solver of 
Roe type \citep{roe1986} and a radiative transfer module based on 
long characteristics and the Feautrier scheme. Multi-group opacities with 
five opacity bands were used. The simulation started 
from a former relaxed simulation of lower spatial resolution, which was
interpolated to the present higher resolution and relaxed for another
30~min physical time. Computations were carried out on two different
machines at the Swiss National Supercomputing Center (CSCS) in Lugano and on a machine at the Geneva Observatory.
The simulation was evolved for 1.5~h physical time after the relaxation
phase and reproduces, with excellent fidelity, the well known granular 
pattern of the solar surface and magnetic bright points in the intergranular 
space if magnetic fields are added. We call this simulation run $\mbox{Roe}_{10}$.

%----------------------------------------------------------------------------
\begingroup
\renewcommand*{\arraystretch}{1.5}
\begin{table}\setlength{\tabcolsep}{4pt}
\centering
\caption{Box and cell sizes, number of opacity bands, $n_\text{opbds}$, and run duration, $t_\text{run}$, for the two simulation runs 
$\mbox{Roe}_{10}$ ({\tt d3t57g45b0roefc})
and $\mbox{Roe}_7$ ({\tt d3t57g44b0}).}
\label{tab:simulation-runs}
\begin{scriptsize}
\begin{tabular}{cccccc}
\hline
\hline
Run & Cell size $[\mbox{km}^3]$ & Box size $[\mbox{Mm}^3]$ & Height range $[\mbox{km}]$ & $n_\text{opbds}$ & $t_\text{run}$ \\
\hline
$\mbox{Roe}_{10}$ & $10\times10\times10$ & $9.6\times9.6\times2.8$ & $-1240<z<1560$ & $5$ & $2.6\,\mbox{h}$ \\
$\mbox{Roe}_7$ & $7\times7\times7$ & $5.6\times5.6\times2.8$ & $-2540<z<260$ & $1$ & $1.5\,\mbox{h}$ \\
\hline
\end{tabular}
\end{scriptsize}
\tablefoot{$z$\,=\,0 refers to the mean optical depth one.}
\end{table}
\endgroup
%----------------------------------------------------------------------------

For the statistical analysis in Sect.\,\ref{sect:pysprop}, we also used the simulation run described in \cite{Freytag2013}, referred to as \verb#d3t57g44b0n025#. It is computed with the same hydrodynamic solver as $\mbox{Roe}_{10}$ but has an 
even finer spatial resolution with $7\,\mbox{km}$ cell-size in all directions; 
we therefore call it $\mbox{Roe}_{7}$ for short. This simulation covers a 
field-of-view of $5600\times5600\,\mbox{km}^{2}$, and was computed with one 
single opacity band (grey opacity) and full boxes were stored every 10~min 
for a time span of 3~h. The two runs are summarized in Table \ref{tab:simulation-runs}.

%----------------------------------------------------------------------------
\begin{figure}
\centering
\includegraphics[width=0.8\columnwidth]{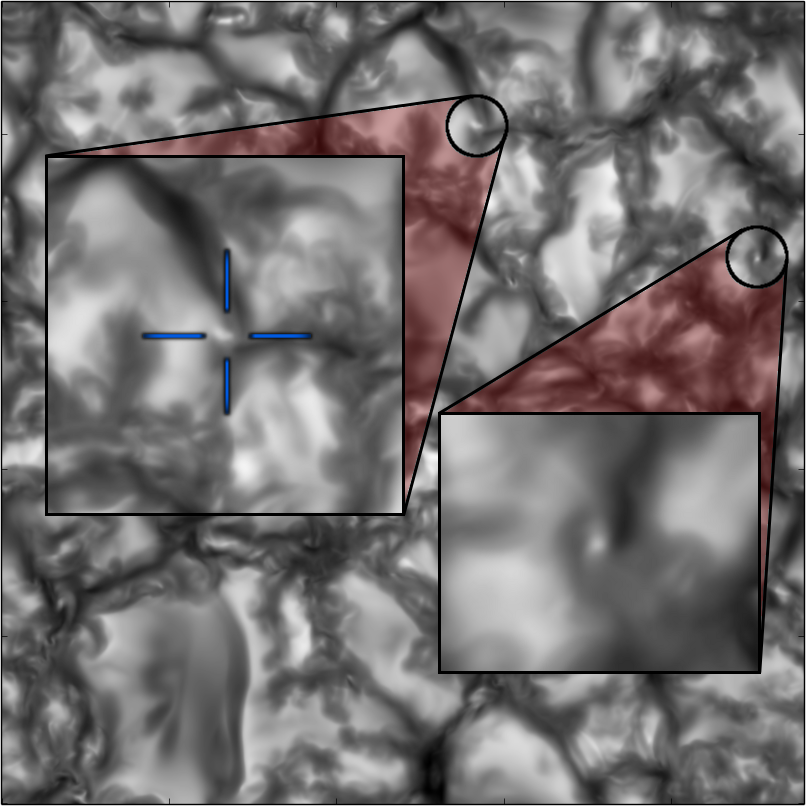}
\caption{Bolometric intensity map of the non-magnetic simulation of
a field of view of $9.6\times 9.6~\mbox{Mm}^2$. The nMBP 
in the close-up on the \emph{left} has a diameter of $80\,\mbox{km}$ (resolved 
by approximately $8\times8$ computational cells). It has an intensity contrast 
of 35\% with respect to the immediate neighbourhood but only 5\% with 
respect to the global average intensity. A second nMBP is visible in
the close-up on the \emph{lower right}. Vertical sections along the blue
lines in the close-up on the \emph{left} are shown in 
Fig.\,\ref{fig:Vertical-slices-through}.
This figure is accompanied by a movie.}
\label{fig:Bolometric-intensity-map}
\end{figure}
%----------------------------------------------------------------------------

Figure\,\ref{fig:Bolometric-intensity-map} shows two nMBPs in one
of the intensity maps taken at $t=240\,\mbox{s}$ after the relaxation
phase of run $\mbox{Roe}_{10}$. Like magnetic bright points, they are located in the intergranular
space but are smaller in size. 
From the accompanying movie one can see 
that nMBPs appear in the intergranular space, often at vertices of 
intergranular lanes and they
move with them or along intergranular lanes in a similar fashion to magnetic
bright points. They often show a rotative proper motion.

\begin{figure*}
\centering
\includegraphics[width=0.98\textwidth]{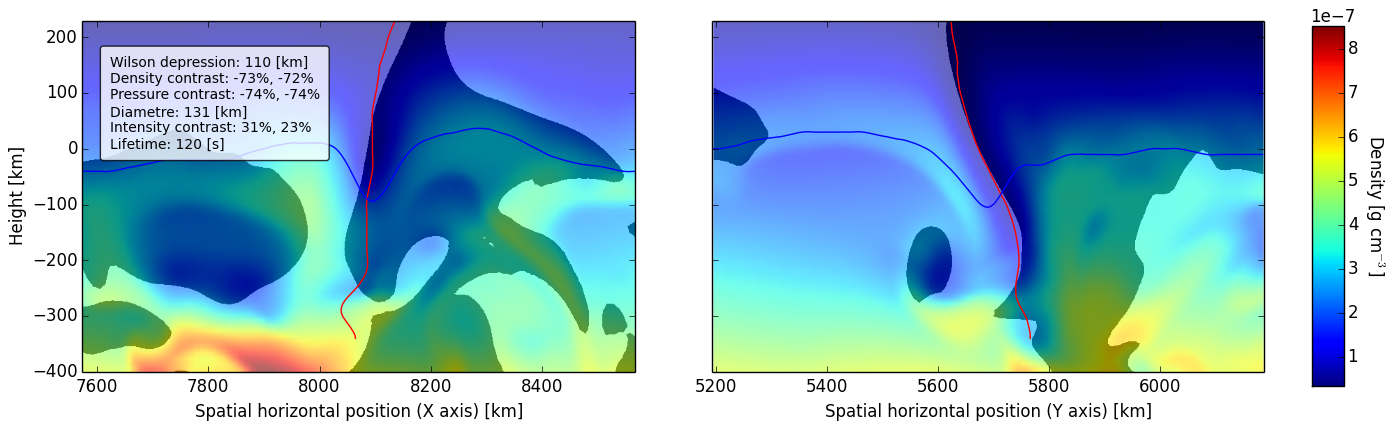}
\caption{Vertical sections through the nMBP shown in 
Fig.\,\ref{fig:Bolometric-intensity-map}. The background colour shows
mass density, where the brightness encodes the sign of the velocity perpendicular 
to the plane of projection: bright is plasma flowing out of the plane and dark
is plasma flowing into it. The red curve indicates the
spine of the nMBP, a ``valley'' in density, found as a sequence of local density 
minima in horizontal planes. The blue curve indicates the Rosseland
optical depth unity. The legend gives intensity, density, and gas pressure 
contrast, firstly with respect to the neighbourhood and secondly with respect
to the global average (at $\tau=1$ for the density and the pressure).}
\label{fig:Vertical-slices-through}
\end{figure*}

Vertical sections of the density in planes along the lines of the  
blue cross indicated in Fig.\,\ref{fig:Bolometric-intensity-map} 
are shown in Fig.\,\ref{fig:Vertical-slices-through}. 
In this figure, the blue curve
indicates the optical depth unity for vertical lines of sight, which
separates the convection zone from the photosphere. One
notices that in the location of the nMBP, there is a funnel of reduced
density (colour scale) that extends from the photosphere into the convection 
zone. The red
curve indicates the local minimum of the density in horizontal planes.
Similar funnels of low density are also found in connection with magnetic 
bright points.
However, in the context of small-scale magnetic flux-tube models
\citep[see, e.g.][]{spruit1976,zwaan1978,steiner+al1986,Fischer2000,Steiner2007}
the pressure gradient caused by this low density region is balanced
by the magnetic pressure of the magnetic field concentration.

%----------------------------------------------------------------------------
%----------------------------------------------------------------------------

The underlying mechanism causing nMBPs is necessarily different because
no magnetic pressure is available. We
can get  information about this mechanism by looking at the Eulerian momentum
equation 
\begin{equation}\label{eq:navier-stokes}
   \frac{\partial\vect{v}}{\partial t}
   +(\vect{v}\cdot\vect{\nabla})\,\vect{v}
   +\frac{1}{\rho}\vect{\nabla}P+\vect{g}=0\,,
\end{equation}
where $\vect{v}$ is the velocity field, $t$ the time, $\rho$ the mass density,
$P$ the gas pressure, and $\vect{g}$ the gravitational acceleration at the
solar surface.
Assuming a stationary field we can neglect the first term.
In a horizontal plane, 
gravity does not play any role, and therefore the pressure gradient can 
only be balanced by a particular shape of the velocity field, which appears 
in the advection term. In Fig.\,\ref{fig:Vertical-slices-through} the sign
of the velocity field is encoded by means of brighter colours (velocity
is going out of the plane of projection) and darker colours (velocity
is going into it). Combining the information provided by the two planes,
we conclude that plasma is swirling around the funnel of low density,
the minimum of which is marked by the red curve.

%----------------------------------------------------------------------------
\begin{figure*}
\centering
\includegraphics[width=0.9\textwidth]{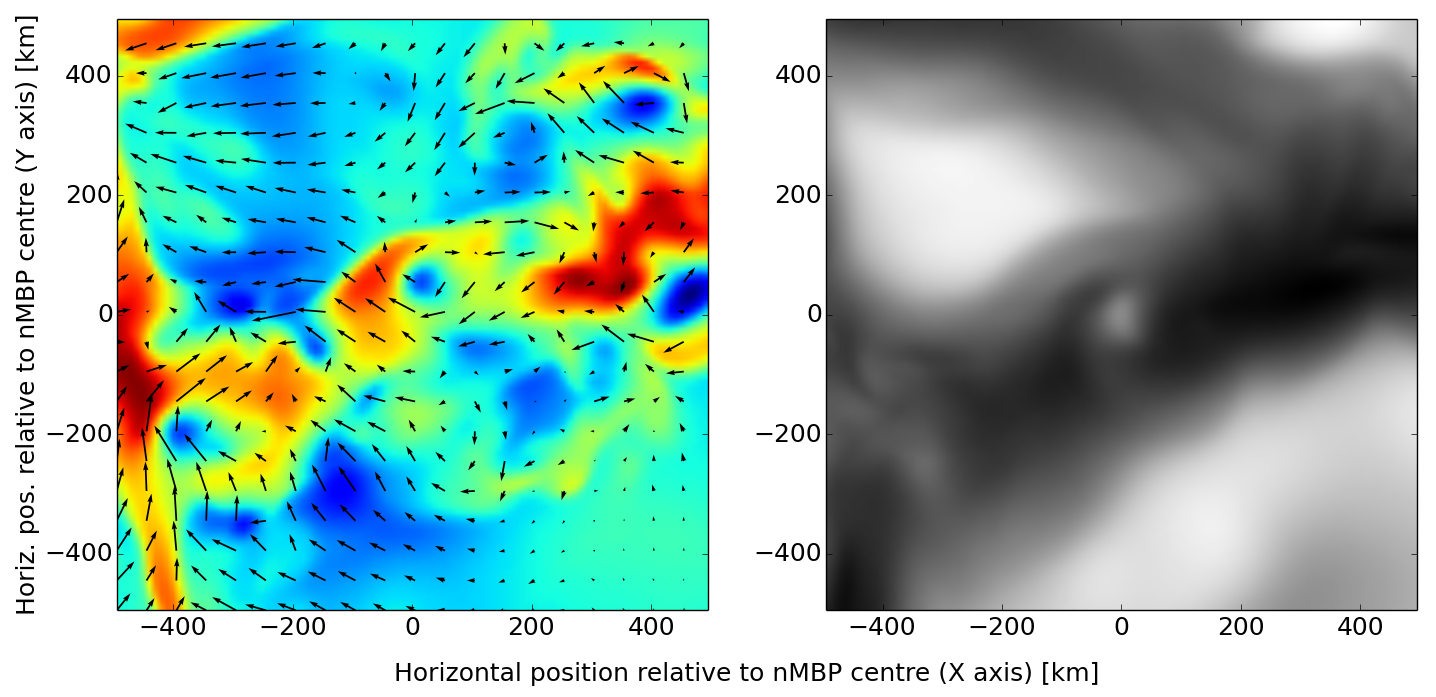}
\caption{\emph{Left:} low-density regions in a sub-surface layer 
$200\,\mbox{km}$ below $\left\langle \tau\right\rangle =1$ sustaining 
swirls. The colour map extends from blue, indicating low density, 
to red, indicating high density, and the arrows represent the plasma 
velocity projected into the horizontal plane. Longest arrows correspond
to a velocity of 9.5\,km\,s$^{-1}$. Such swirling low density regions are 
much more abundant than nMBPs but most of them do not
extend into the photosphere and therefore do not produce nMBPs. 
\emph{Right:} 
corresponding bolometric intensity map. Of the several
low density swirling regions in the left panel only the one
at $(x,y) = (0,0)$ produces a nMBP.}
\label{fig:swirls}
\end{figure*}
%----------------------------------------------------------------------------

This swirling explains the stability of nMBPs only qualitatively,
but it also provides insight into their formation. The plasma that rises to
the photosphere by convection cools off, becomes denser, and subsequently 
sinks back into the convection zone in the intergranular lanes, 
which are 
smaller than the granules. Any small initial angular momentum
in the plasma will lead, when downflows contract, to the generation
of swirls. This effect has already been described by \cite{Nordlund1985}
in a slightly different context. There, it was referred as the ``bath-tub'' 
effect and appeared in the sub-surface layers at locations of
downdrafting plumes. In our simulations we also find many locations
with low densities in the sub-surface layers and corresponding swirls
(see Fig.\,\ref{fig:swirls}), but in general those swirls do not lead to
bright points. In fact, bright points only appear if the low density
funnel extends spatially from the convection zone to a significant
height into the photosphere.
Such an extended region of low density leads automatically to a depression
of the unit optical depth isosurface 
(blue contour in Fig.\,\ref{fig:Vertical-slices-through}),
for which we will borrow the term ``Wilson depression'' in analogy
to the observed depression in sunspots.

%
%%%%%%%%%%%%%%%%%%%%%%%%%%%%%%%%%%%%%%%%%%%%%%%%%%%%%%%%%%%%%%%%%%%%%%%%%%%%%
%
\section{From a nMBP toy-model to a recognition algorithm}
\label{sect:toy}
We now come back to the Eulerian momentum equation, Eq.\,\eqref{eq:navier-stokes}.
For establishing a toy model, we impose the following conditions, which derive 
from the properties of typical nMBPs, as the one shown in 
Figs.\,\ref{fig:Bolometric-intensity-map}--\ref{fig:swirls}:
\begin{enumerate}
   \item nMBPs are long-lived and stable so that the velocity field can
         be considered stationary;
   \item they have cylindrical symmetry;
   \item their velocity field has a non-vanishing azimuthal component;
   \item they extend in the vertical direction and their shape does not 
         depend on depth.
   \end{enumerate}
Because of the second assumption, the Euler momentum equation should
be rewritten in cylindrical coordinates. The advection term is then
given by
\begin{equation*}
   (\vect{v}\cdot\vect{\nabla})\vect{v}=
   \left[(\vect{v}\cdot\vect{\nabla})v_{r}-
   \frac{v_{\theta}^{2}}{r}\right]\vect{\hat{r}}+
   \left[(\vect{v}\cdot\vect{\nabla})v_{\theta}+
   \frac{v_{\theta}v_{r}\vphantom{v_{\theta}^{2}}}{r}\right]
   \vect{\hat{\theta}}+(\vect{v}\cdot\vect{\nabla})v_{z}\vect{\hat{z}}\,,
\end{equation*}
where the directional derivative is 
\begin{equation*}
   \vect{v}\cdot\vect{\nabla}=
   v_{r}\partial_{r}+\frac{v_{\theta}}{r}\partial_{\theta}
   +v_{z}\partial_{z}\,.
\end{equation*}
The simplest field satisfying the conditions 1--4 is
\begin{equation*}
   \vect{v}=v_{\theta}\left(r\right)\vect{\hat{\theta}}\,.
\end{equation*}
Because of cylindrical symmetry, the pressure gradient is provided
by $\left(\partial_{r}P\right)\vect{\hat{r}}+\left(\partial_{z}P\right)\vect{\hat{z}}$.
Inserting the velocity field into the Euler momentum equation and
projecting it into the horizontal plane leads to the trivial statement
that the pressure gradient is in fact provided by the centripetal force
\begin{equation}\label{eq:cp}
   \frac{1}{\rho}\frac{\partial P}{\partial r}=\frac{v_{\theta}^{2}}{r}\,.
\end{equation}

When comparing the typical nMBPs of Figs.\,\ref{fig:Bolometric-intensity-map}
and \ref{fig:Vertical-slices-through} to this model, it immediately turns out
that assumptions 2 and 4 are only qualitatively fulfilled.
The first assumption is less straight-forward; on the one hand, the
dynamical timescale for nMBPs is
\begin{equation*}
   \tau=\frac{v}{\left\Vert \displaystyle
   \frac{\partial\vect v}{\partial t}\right\Vert }=
   \frac{v}{\left\Vert (\vect{v}\cdot\vect{\nabla})\,\vect{v}+
   \frac{1}{\rho}\vect{\nabla}P+\vect{g}\right \Vert}\,,
\end{equation*}
and this timescale is in the order of seconds ($6\pm2$~s in
a $1000\times1000\,\mbox{km}^{2}$ neighbourhood around the nMBP of
Fig.\,\ref{fig:Bolometric-intensity-map}, on the $\tau$\,=\,1 isosurface).
On the other hand, the simulations show nMBPs to have a typical lifespan
in the order of the granulation timescale. The difference between
the two timescales can in fact be expected when inspecting the simulations,
because even though nMBPs exist for up to five minutes, they are moving,
often even rotating, and their shape is continuously changing.
The model could be refined by rewriting the velocity field as
$\vect{v}\left(x,t\right)=\vect{v}_{0}\left(t\right)+
\vect{v}^{\prime}\left(x\right)$,
where $\vect{v}_{0}$ is the bulk velocity of the nMBP and $\vect{v}^{\prime}$
is the stationary field that we have considered until now. We then
get the new Euler momentum equation
\begin{equation*}
  \frac{1}{\rho}\frac{\partial P}{\partial r}=
  \frac{v_{\theta}^{2}}{r}+\frac{\partial\vect{v}_{0}}{\partial t}\,,
\end{equation*}
where, in general, $\vect{v}_{0}$ should also be a non-constant 
coordinate-dependent
field. These considerations lead to the conclusion that no simple
model can fully reproduce nMBPs' behaviour. Nevertheless, the present
toy model gives important insight into nMBPs and provides three correlated
characteristics of them:
       (i) the presence of a pressure gradient, leading to a funnel of reduced density 
       by virtue of the equation of state and therefore to a depression of the 
       $\tau$\,=\,1 isosurface;
       (ii) the presence of swirling motion and with it vorticity in the velocity field;
       (iii) a local intensity contrast due to a temperature contrast at the location of
       the depressed $\tau$\,=\,1 isosurface.
Based on this information, we have developed an algorithm by constructing
an indicator (a growing function of pressure gradient, vorticity, and
intensity contrast), and subsequent search of maxima
of this indicator. More precisely, the steps are:
\begin{enumerate}
\item Compute the indicator at $\tau=1$.
      We define $h_{\tau=1}$ as the depth at which $\tau=1$, 
      and $\omega_{v,\,z}\equiv\left(\nabla\times\vect{v}\right)_{z,\,\tau=1}$
      the vertical component of the vorticity, and $T_{\tau=1}$ the temperature,
      all evaluated over the entire surface of $\tau=1$. Then, we define $\N$
      as a normalization operator applied to the array of values of each
      physical quantity, which linearly maps each value to a new value
      in the $\left[0,1\right]$ interval. More precisely, given an array
      $A_{ij}$,
      \begin{equation*}
        \N\left(A\right)_{ij}\equiv
        \frac{A_{ij}-\min_{m,n}\left(A_{mn}\right)}
        {\max_{m,n}\left(A_{mn}\right)-\min_{m,n}\left(A_{mn}\right)}\,,
      \end{equation*}
      and the indicator function is then constructed empirically as
      \begin{equation*}
        \I=\sqrt{\N\left(T_{\tau=1}\right)^{2}+
        \N\left(-h_{\tau=1}\right)^{2}+
        \N\left(\left|\omega_{v,\,z}\right|\right)^{2}}\,.
      \end{equation*}
\item Find the location of maxima of the indicator using a maximum filter 
      (with sliding window of size $3\times3$).
\item For each local maximum of the indicator, search in the intensity map 
      for the closest intensity maximum. This defines the location of a 
      candidate nMBP.
\item Select granules (which we define as regions where the vertical velocity
      is positive at $\tau=1$) and intergranular lanes (complementary region).
\item Select the intergranular, local neighbourhood in the intensity map for 
      every candidate nMBP within an area of $100\times 100$ computational 
      cells centred 
      on the nMBP. The boundary between the nMBP and the local neighbourhood
      is consistently defined to separate the region of intensity
      $I>\frac{1}{2}\left(I_{\mathrm{centre}}+I_{\mathrm{out}}\right)$,
      the nMBP, from the region of intensity $I<\frac{1}{2}
      \left(I_{\mathrm{centre}}+I_{\mathrm{out}}\right)$,
      the neighbourhood. $I_{\mathrm{centre}}$ is the maximal intensity
      in the nMBP region and $I_{\mathrm{out}}$ is the average intensity
      in the neighbourhood.
\item Select the spine of the nMBP in three-dimensional space. To keep the 
      computational demand low, we first extract from the full computational 
      domain a smaller box of $100\times100\times100$ computational cells 
      around its centre and the $\left\langle \tau\right\rangle =1$ surface. 
      Starting at the depth of $\tau=1$ in the centre of the nMBP and from
      there, layer by layer upwards and downwards, we look for local minima 
      in the density. Exactly at $\tau=1$ we select the local density minimum
      closest to the approximate location of the nMBP as determined in step 3. 
      For the adjacent layers, the selected local minimum is the one that is
      closest to that of the previous layer. 
\item Apply step 5 to scalar physical quantities other than the
      bolometric intensity, on either the $z$\,=\,0 plane or the
      $\tau$\,=\,1 isosurface. Evaluate the local and global contrast
      of the corresponding scalar quantity (density, gas pressure, and 
      temperature).
\end{enumerate}
This algorithm managed very well to recognize all the nMBPs that we
had previously selected by eye in an arbitrary snapshot. We then
applied this algorithm to the entire time sequence of the simulation, producing
similar plots as shown in Figs.\,\ref{fig:Vertical-slices-through}
and \ref{fig:swirls}. We have examined all the selected nMBPs visually
and discarded those that were not properly selected (i.e. ``twin'' 
nMBPs that were very close to each other).

%
%%%%%%%%%%%%%%%%%%%%%%%%%%%%%%%%%%%%%%%%%%%%%%%%%%%%%%%%%%%%%%%%%%%%%%%%%%%%%
%
\section{Physical properties of nMBPs}
\label{sect:pysprop}
We investigate the following physical properties of nMBPs:
\begin{description}
\item [\emph{Equivalent diameter $\D$:}] computed from the area $A$ of
   a nMBP as determined in step 5 of the algorithm given in the previous
   section. Thus, $D\equiv 2\sqrt{{A}/{\pi}}$.
\item [\emph{Wilson-depression $\WD$:}] difference in height between the
   deepest point on the $\tau$\,=\,1 isosurface of a nMBP and the average height
   of this surface in the neighbourhood of the nMBP as determined in step 5 of 
   the algorithm. 
\item [\emph{Intensity contrast $\Ic$:}] computed locally and globally from
   $\Ic \equiv {(I_{p}-\left\langle I\right\rangle)}/{\left\langle I\right\rangle }$,
   where $I_{p}$ is the average bolometric intensity of the nMBP and 
   $\left\langle I\right\rangle $ is either
   the bolometric intensity averaged over the entire intensity map (global) or in
   the neighbourhood (local) of the nMBP as determined in step 5 of the algorithm.
\item [\emph{Mass-density contrast $\Rhoc$:}] computed locally and globally
   on the horizontal plane corresponding to $\left\langle \tau\right\rangle =1$
   (i.e., $z=0$) and on the $\tau$\,=\,1 isosurface, respectively. 
   As determined in step 7 of the algorithm,
   $\Rhoc\equiv {(\rho_{0}-\left\langle \rho\right\rangle)}/
   {\left\langle \rho\right\rangle }$,
   where $\rho_{0}$ is the minimal mass density inside the nMBP at the
   given surface and $\left\langle \rho\right\rangle $ is either the
   density averaged over the entire plane or in the neighbourhood at
   the given surface.
\item [\emph{Pressure contrast $\Pc$:}] defined in the same way as the mass
   density contrast, with $p_{0}$ taken at the same location as
   $\rho_{0}$.
\item [\emph{Temperature contrast $\Tc$:}] defined in the same way as the
   mass density and the pressure contrast, with $T_{0}$ taken at the
   same location as $\rho_{0}$.
\item [\emph{Maximum vertical velocity $\VvMax$:}] defined in the horizontal 
   plane $z=0$ (corresponding to $\left\langle \tau\right\rangle =1$), inside the 
   nMBP.
\item [\emph{Maximum horizontal velocity $\VhMax$:}] also defined in the horizontal
   plane $z=0$ (corresponding to $\left\langle \tau\right\rangle =1$), inside
   the nMBP.
\item [\emph{Vorticity $\Vort$:}] vertical component of the vorticity, defined
   in the horizontal plane $z=0$ at the same location as $\rho_{0}$.
\item [\emph{Lifetime:}] derived from a few nMBPs visually tracked
   on bolometric intensity maps of $\mbox{Roe}_{10}$. 
\end{description}

The nMBPs that enter the following statistics have been extracted from 
the two distinctly different magnetic field-free simulation runs
$\mbox{Roe}_{10}$ and $\mbox{Roe}_7$ (see Table\,\ref{tab:simulation-runs}). 
The mean values and corresponding standard deviations of the physical 
quantities listed
above (with the exception of the lifetime) are given in 
Table\,\ref{tab:physical-quantities} for both simulation runs and
(depending on the quantity)
with respect to the plane at $z=0$ and/or the $\tau$\,=\,1 surface and
with respect to the local and global neighbourhood. 
For the lifetime, no automatic tracking of nMBPs was done 
because the cadence of 240 s of the full box data is not high enough. 
On the other hand, bolometric intensity maps are available at a high 
cadence of 10~s,
but the tracking of  nMBPs from the intensity maps alone would require
involved pattern recognition techniques and criterions for the 
appearance and disappearance of nMBPs. Therefore we relied on a derivation 
from visually tracked nMBPs on the bolometric intensity maps.
Qualitatively, we found bright points existing for a duration of 30~s up 
to the granular time-scale of a few min.

%----------------------------------------------------------------------------
\begingroup
\renewcommand*{\arraystretch}{1.5}
\begin{table}\setlength{\tabcolsep}{4pt}
\centering
\caption{Physical quantities of nMBPs of two different simulations, 
Roe$_{10}$ with a grid size of $10\,\mbox{km}$ 
in every spatial direction and Roe$_{7}$ with a grid size 
of $7\,\mbox{km}$ in every direction. $\mu$ is the mean value,
and $\sigma$ the standard deviation. See text for explanation of the
physical quantities.}
\label{tab:physical-quantities}
\begin{tabular}{c|c|c|c|c|c|c}
\cline{4-7} 
\hline
\hline
\multicolumn{3}{c|}{Simulation run} & \multicolumn{2}{c|}{Roe$_{10}$} & \multicolumn{2}{c}{Roe$_{7}$}\tabularnewline
\cline{4-7} 
\hline
\multicolumn{3}{c|}{Mean and standard deviation} & $\mu$ & $\sigma$ & $\mu$ & $\sigma$\tabularnewline
\hline 
\multicolumn{3}{c|}{$\D$ {[}km{]}} & 80 & 21 & 62 & 15\tabularnewline
\multicolumn{3}{c|}{$\WD$ {[}km{]}} & 103 & 32 & 83 & 34\tabularnewline
\hline 
\multirow{2}{*}{$\Ic$} & \multicolumn{2}{c|}{Local} & 21\% & 10\% & 19\% & 10\%\tabularnewline
 & \multicolumn{2}{c|}{Global} & 2\% & 9\% & -1\% & 12\%\tabularnewline
\hline 
\multirow{4}{*}{$\Rhoc$} & \multirow{2}{*}{At $z=0$} & Local & -59\% & 10\% & -54\% & 12\%\tabularnewline
 &  & Global & -58\% & 10\% & -57\% & 11\%\tabularnewline
\cline{2-7} 
 & \multirow{2}{*}{At $\tau=1$} & Local & -43\% & 14\% & -39\% & 19\%\tabularnewline
 &  & Global & -41\% & 15\% & -40\% & 17\%\tabularnewline
\hline 
\multirow{4}{*}{$\Pc$} & \multirow{2}{*}{At $z=0$} & Local & -59\% & 10\% & -54\% & 12\%\tabularnewline
 &  & Global & -61\% & 10\% & -60\% & 12\%\tabularnewline
\cline{2-7} 
 & \multirow{2}{*}{At $\tau=1$} & Local & -38\% & 15\% & -35\% & 19\%\tabularnewline
 &  & Global & -38\% & 16\% & -37\% & 18\%\tabularnewline
\hline 
\multirow{4}{*}{$\Tc$} & \multirow{2}{*}{At $z=0$} & Local & -0.7\% & 2\% & -2\% & 4\%\tabularnewline
 &  & Global & -9\% & 4\% & -8\% & 8\%\tabularnewline
\cline{2-7} 
 & \multirow{2}{*}{At $\tau=1$} & Local & 8\% & 3\% & 6\% & 3\%\tabularnewline
 &  & Global & 5\% & 4\% & 5\% & 4\%\tabularnewline
\hline 
\multicolumn{3}{c|}{$\VvMax$ {[}km\,s$^{-1}${]}} & $-6.9$ & 2.9 & -5.5 & 4.5\tabularnewline
\multicolumn{3}{c|}{$\VhMax$ {[}km\,s$^{-1}${]}} & $9.5$ & 1.3 & 9.2 & 1.1\tabularnewline
\multicolumn{3}{c|}{\tiny $\sqrt{R\,T_{\mathrm{eff}}\left[1-\left(1+C_{\rho}\right)\left(1+C_{T}\right) \right]}$} & 4.8 & 0.40 & 4.6 & 0.57\tabularnewline
\multicolumn{3}{c|}{$\left|\Vort\right|$ {[}$\mbox{s}^{-1}${]}} & 0.42 & 0.12 & 0.35 & 0.14\tabularnewline
\multicolumn{3}{c|}{$\Vort$ {[}$\mbox{s}^{-1}${]}} & 0.02 & 0.44 & -0.02 & 0.38\tabularnewline
\multicolumn{3}{c|}{$\sqrt[4]{1+\Ic}-1\,\left(\mbox{at}\,\tau=1,\ \mbox{local}\right)$} & 5\% & 2\% & 4\% & 2\%\tabularnewline
\hline 
\end{tabular}
\tablefoot{See text for explanation of the physical quantities.}
\end{table}
\endgroup
%----------------------------------------------------------------------------

\begin{figure}
\centering
\includegraphics[width=1.0\columnwidth]{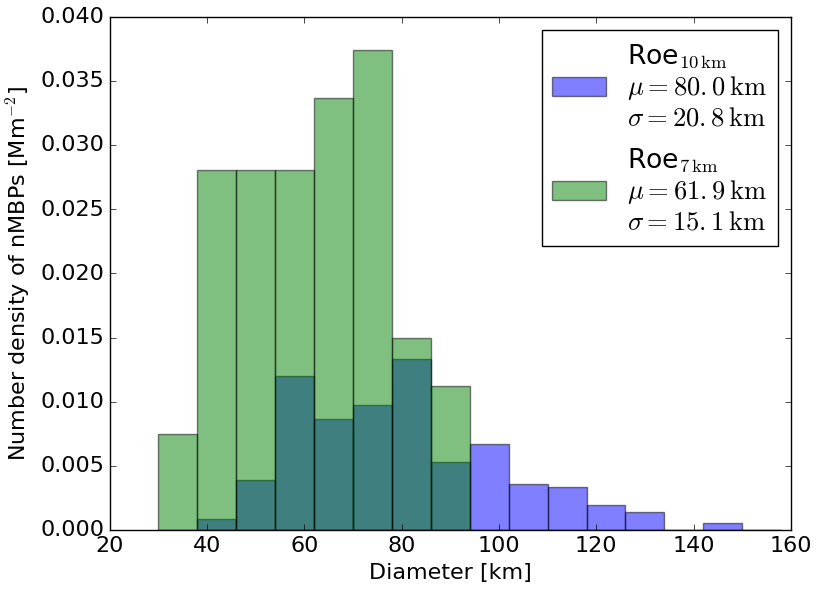}
\caption{Size distribution of nMBPs, normalized to the average number 
density in the corresponding simulation.}
\label{fig:size-distribution}
\end{figure}

\begin{figure}
\centering
\includegraphics[width=1.0\columnwidth]{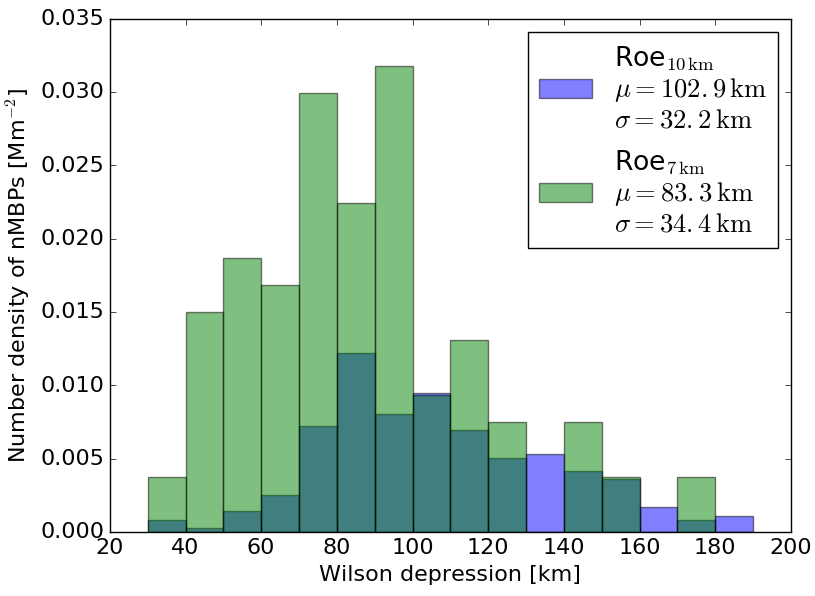}
\caption{Distribution of the Wilson-depression of nMBPs, normalized to  
the average number density in the corresponding simulation.}
\label{fig:wd-distribution}
\end{figure}

For the first two entries in Table\,\ref{tab:physical-quantities},
the equivalent diameter $D$ and the Wilson depression $\WD$,
the corresponding histograms are given in Figs.\,\ref{fig:size-distribution}
and \ref{fig:wd-distribution}, respectively.
From Fig.\,\ref{fig:size-distribution} one readily sees that the 
diameters of nMBPs are distinctly smaller than those of magnetic 
bright points, which range from 100\,km to 300\,km according to
\citet{wiehr+al2004}. The median value for the diameter is 78\,km 
and 63\,km for $\mbox{Roe}_{10}$ and $\mbox{Roe}_{7}$, respectively,
while \citet{wiehr+al2004} obtain 160\,km for the most frequent 
diameter of magnetic bright points.
In Fig.\,\ref{fig:size-distribution}, one also observes that the 
distribution of diameters of $\mbox{Roe}_{7}$ is rather sharply limited 
at $30\,\mbox{km}$ and $95\,\mbox{km}$, which could suggest that nMBPs are about to be resolved in this simulation. 
Lowering the resolution can be expected to spread this distribution, 
which would then explain the apparition of wings in the size distribution 
of $\mbox{Roe}_{10}$. But more likely, the low end of the distribution
is given by the limited spatial resolution of the simulations. On the other 
hand, we cannot offer a plausible explanation for the extended wide wing 
to larger diameters in the histogram of $\mbox{Roe}_{10}$.

Besides the physical quantities that appear on 
Table\,\ref{tab:physical-quantities},
another interesting quantity is the number of nMBPs per unit area,
$n_{10}$ and $n_{7}$. We find $n_{10}=0.0712\,\mbox{Mm}^{-2}$
and $n_{7}=0.189\,\mbox{Mm}^{-2}$, which indicates that at higher
spatial resolution, we observe more than twice as many bright points than
at lower resolution. The histograms of 
Figs.\,\ref{fig:size-distribution} and \ref{fig:wd-distribution}
are normalized to these respective number densities.
The histogram of the equivalent diameter shown in 
Fig.\,\ref{fig:size-distribution}
suggests that this surplus is due to a larger number of small bright 
points detected at higher spatial resolution. However, there still 
remains a difference of a factor of two between number densities
in the two simulations when restricting statistical analyses to  nMBPs 
larger than $60\,\mbox{km}$ in diameter; a value that is significantly greater
than the spatial resolution of both simulations. This suggests
that the high resolution simulation not only harbours more small 
and tiny nMBPs but also greatly favours the formation of large nMBPs.

%----------------------------------------------------------------------------
%----------------------------------------------------------------------------

%----------------------------------------------------------------------------
%----------------------------------------------------------------------------

The distribution of the Wilson depression, given in Fig.\,\ref{fig:wd-distribution},
also shows an interesting trend: in the higher resolution model $\mbox{Roe}_{7}$
, it is globally shifted towards smaller depths compared to the distribution
from $\mbox{Roe}_{10}$. An intuitive explanation for this fact would
be a correlation between size and Wilson depression, which, however, does 
not exist. There are a number of numerical parameters that differ between
simulations $\mbox{Roe}_{10}$ and $\mbox{Roe}_{7}$ but numerous 
test runs confirmed that none of these
seem to substantially influence the Wilson depression. We therefore
found no convincing explanation why simulation $\mbox{Roe}_{7}$ shows
nMBPs to have a mean Wilson depression which is approximately 20\,km less than that
of simulation $\mbox{Roe}_{10}$. In any case, the Wilson depression of
nMBPs of respectively 103\,km and 83\,km for $\mbox{Roe}_{10}$ and 
$\mbox{Roe}_{7}$ is again clearly smaller than corresponding values 
for magnetic bright points of, typically, 150\,km (Salhab et al., in prep.).

In Table\,\ref{tab:physical-quantities}, one can see that the average
values of contrasts of intensity, mass density, pressure, and temperature
do not differ substantially between the two simulations. At first, this seems to be at 
odds with the fact that the Wilson depression is larger in $\mbox{Roe}_{10}$
compared to $\mbox{Roe}_{7}$. Thinking of a plane-parallel, exponentially-stratified, hydrostatic atmosphere one would expect the density contrast 
at $\tau=1$ to also be larger in $\mbox{Roe}_{10}$ compared to 
$\mbox{Roe}_{7}$. However, this is only partially the case. The reason
is that close to $\tau=1$, the density runs rather constant with 
height due to the onset of radiative loss and a corresponding sharp drop 
in temperature (this is the quasi density reversal that causes the 
Rayleigh-Taylor instability that drives the convection). Furthermore, 
the funnel-shaped structure of the nMBPs accentuates this behavior; smaller 
radii in deeper layers go together with faster swirling motion that can
sustain larger gradients in mass density and pressure, which  
counteracts stratification of the atmosphere inside the nMBP. The absence 
of exponential stratification then keeps the density contrast insensitive 
to the Wilson depression. 

The intensity contrast (third entry in Table\,\ref{tab:physical-quantities})
is a quantity that can be directly observed.
As is the case for the contrast of temperature at the $\tau=1$ isosurface,
its value drastically depends on whether it is evaluated locally (with
respect to the local neighbourhood) or globally (with respect to the
mean intensity/temperature). This is due to the fact that nMBPs are
formed within the dark intergranular space and are, compared to granules,
not so bright. Some of the nMBPs are even darker than the average
intensity over the whole bolometric map, but they are still conspicuous
objects within the dark intergranular lanes 
\citep[see also][for the same effect in the context of 
magnetic bright points]{title+berger1996}.
The contrast of the bolometric intensity of nMBPs in the order of 
20\% (local) and 0\% (global) is again distinctively smaller than the 
measured global continuum contrast of magnetic bright points at 588\,nm 
of 10\% to 15\% by \citet{wiehr+al2004} and at
525\,nm of 37\% (local) and 11\% (global) by \citet{riethmueller+al2010}.
We ascribe this difference to the difference in the formation
of magnetic bright points and nMBPs. Magnetic flux tubes are prone
to the convective collapse instability \citep{spruit1979}, which leads 
to a high degree of evacuation and with it to a large Wilson depression 
and intensity contrast. No such instability and evacuation takes place 
in the case of nMBPs.
Like \citet{wiehr+al2004} for magnetic bright points, we do not find
a correlation between local intensity contrast and effective diameter
of nMBPs.

\begin{figure*}[t]
\centering
   \includegraphics[width=0.9\textwidth]{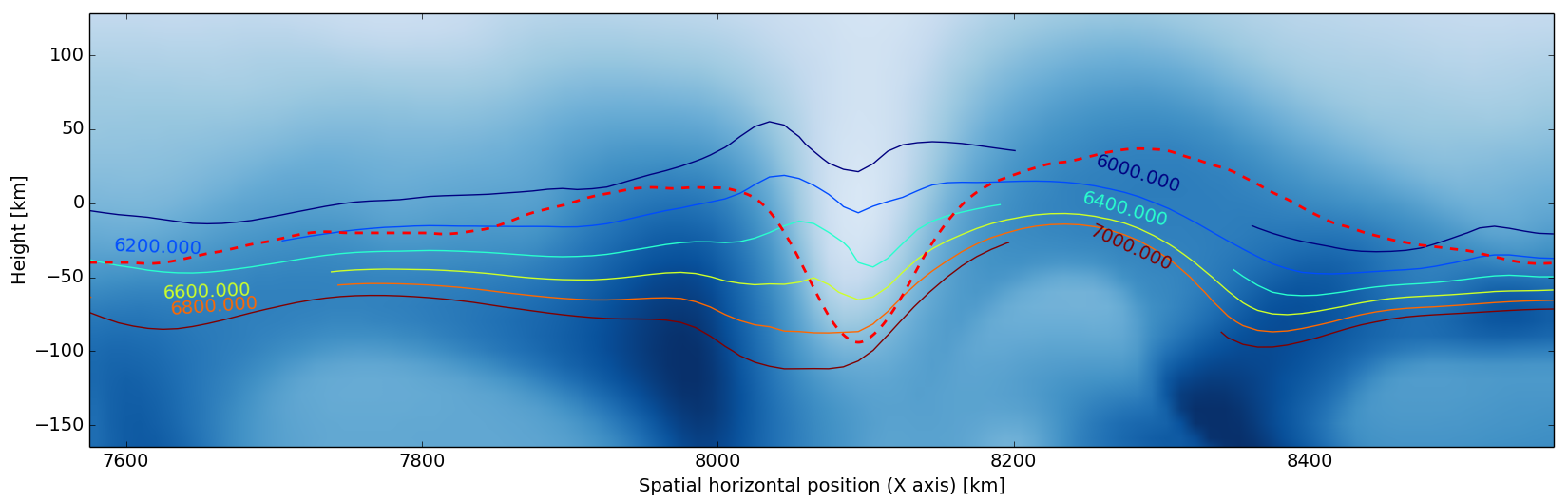}
\caption{Vertical section through a typical nMBP showing isotherms 
(labeled coloured curves) and the mass density plotted in the background 
(light is low density and dark blue is high density). 
The $\tau$\,=\,1 isosurface is the dashed red curve. 
This surface intersects the isotherms, reaching to high
temperatures where it dips deep down at the location of the nMBP
at $x\approx 8100$\,km.}
\label{fig:isotherms}
\end{figure*}

The temperature contrast on the isosurface of unit optical depth is
positive as can be seen from Table\,\ref{tab:physical-quantities}. 
This can be seen in Fig.\,\ref{fig:isotherms}, which
shows that isotherms intersect the $\tau$\,=\,1 isosurface, similarly to 
magnetic bright points \citep[see, e.g. Fig.\,2 of][]{Steiner2007}. 
It is interesting to note that isotherms 
generally also have a depression at the location of the nMBP funnel, 
although smaller than the depression of optical depth unity. This
can be expected because of the swirling downdraft of cool 
photospheric material at locations of nMBPs. The temperature contrast on a horizontal plane of 
constant geometrical depth can therefore be negative,
although it is always positive on the $\tau$\,=\,1 isosurface, which is also
the reason for the brightness of nMBPs. Pressure and density contrasts have
very similar values, which is expected because both quantities are
related by the equation of state, while the temperature contrast is
relatively low. The nMBPs form in swirling downdrafts in the
intergranular space. The maximal vertical velocity ranging from 5.5 to 7\,km\,s$^{-1}$
approaches sonic speed.

%----------------------------------------------------------------------------
%----------------------------------------------------------------------------

The vertical vortex event described by \citet{moll+al2011} has a
diameter of 80\,km, a Wilson depression of 110\,km, an intensity 
contrast of 0.24, and global density and pressure contrasts at 
$z=0$ of $-0.64$. This is one example observed in a simulation
with a grid size of $4\times 4\times 4$\,km$^3$ as obtained with the
MURaM code \citep{voegler+al2005}. These values are in close
agreement with the values and standard deviations
given in Table\,\ref{tab:physical-quantities} suggesting that this vortex
event is of the same class of objects described herein. 
Correspondingly, we can consider nMBPs to be manifestations of 
vertically extending vortex tubes, which themselves are the manifestation
of the turbulent nature of the near surface convection.

Regarding the horizontal, azimuthal velocities in nMBPs, the toy model in 
Sect.\,\ref{sect:toy} provides an order of magnitude estimate. For this, 
we consider the relations
\begin{alignat*}{4}
   \frac{P_{\mathrm{ext}}-P_{\mathrm{int}}}{\rho_{\mathrm{ext}}} 
   &\approx  v_{\theta}^{2}\,,\quad
   &\frac{\rho_{\mathrm{int}}-\rho_{\mathrm{ext}}}{\rho_{\mathrm{ext}}} 
   &\equiv C_{\mathrm{\ensuremath{\rho}}}\,,\quad
   &\frac{T_{\mathrm{int}}-T_{\mathrm{ext}}}{T_{\mathrm{ext}}} 
   &\equiv  C_{T}\,,\\
   \frac{P_{\mathrm{ext}}}{\rho_{\mathrm{ext}}} 
   &\approx R_{\mathrm{s}}\,T_{\mathrm{ext}}\,,
   &T_{\mathrm{ext}} 
   &\approx  T_{\mathrm{eff}}\,,
   &&
\end{alignat*}
where $P$, $\rho$, $C_{\rho}$, $T$, $C_{T}$, $T_{\mathrm{eff}}$
and $v_{\theta}$ stand for gas pressure, mass density,
density contrast, temperature, temperature contrast, effective temperature,
and tangential velocity respectively, and $R_{\mathrm{s}}$ is the 
specific gas constant for which we choose the mean molecular weight
$\mu = 1.224$. The indices ``int'' and ``ext'' refer to the centre
of the nMBP funnel and to its close surrounding, respectively, always
in the horizontal plane of $\left\langle \tau\right\rangle =1$. 
The first equation is derived from Eq.~(\ref{eq:cp}). 
One can then express the tangential velocity
as a function of density contrast, temperature contrast, and effective
temperature as
\begin{equation*}
  v_{\theta}=\sqrt{R_{\mathrm{s}}\,T_{\mathrm{eff}}
  \left[1-(1+C_{\rho})(1+C_{T})\right]}
  \approx\sqrt{\frac{R_{\mathrm{s}}\,T_{\mathrm{eff}}}{2}}=4.4\,
  \mbox{km}\,\mbox{s}^{-1}\,.
\end{equation*}
Here, we used $C_{\rho}\approx -0.5$
and $C_{T}\approx 0$ from Table\,\ref{tab:physical-quantities},
referring to the horizontal plane $\left\langle \tau\right\rangle =1$. 
This value is approximately half the actual maximal horizontal velocity 
of nMBPs in the simulations, which is, close to nMBP centres, around 
9--9.5\,km\,s$^{-1}$
according to Table\,\ref{tab:physical-quantities}. 

As expected, there are as many bright points rotating clockwise 
as there are rotating anti-clockwise. We verified that the average value of
the vertical component of the vorticity $\left\langle \Vort\right\rangle $
is indeed close to zero.

Using both the Eddington-Barbier relation and the Stefan-Boltzmann law, 
the temperature contrast at $\tau=1$ and the intensity contrast 
are related as
\begin{equation*}
 1+C_{I}\approx\left(1+C_{T}\right)^{4}\,.
\end{equation*}
It is verified in the simulations that
\begin{equation*}
  \left\langle \frac{1+C_{I}}
  {\left(1+C_{T}\right)^{4}}\right\rangle =1.01\pm0.06,
\end{equation*}
and the Pearson cross-correlation coefficient for these two contrasts
is given by $\rho=0.49$ ($\mbox{Roe}_{10}$ simulation)
and $\rho=0.35$ ($\mbox{Roe}_{7}$ simulation).

%
%%%%%%%%%%%%%%%%%%%%%%%%%%%%%%%%%%%%%%%%%%%%%%%%%%%%%%%%%%%%%%%%%%%%%%%%%%%%%
%
\section{nMBPs in intensity maps of virtual observations}
\label{sect:synthobs}
The kind of nMBPs described here have not been observed 
with currently operating solar telescopes, presumably because of their small 
size and relatively low intensity contrast. Nonmagnetic photospheric bright 
points have been reported to exist by \citet{berger+title2001} and by 
\citet{langhans+al2002}. However, those bright points exist on the edges 
of certain bright, rapidly expanding granules while nMBPs are 
located within intergranular lanes.
The objects reported by \citet{berger+title2001} and by \citet{langhans+al2002} are probably identical 
with the bright grains that often appear with the development of granular lanes
\citep{yurchyshyn+al2011}.

%----------------------------------------------------------------------------
\begin{figure*}[t]
\centering
   \includegraphics[width=0.47\textwidth]{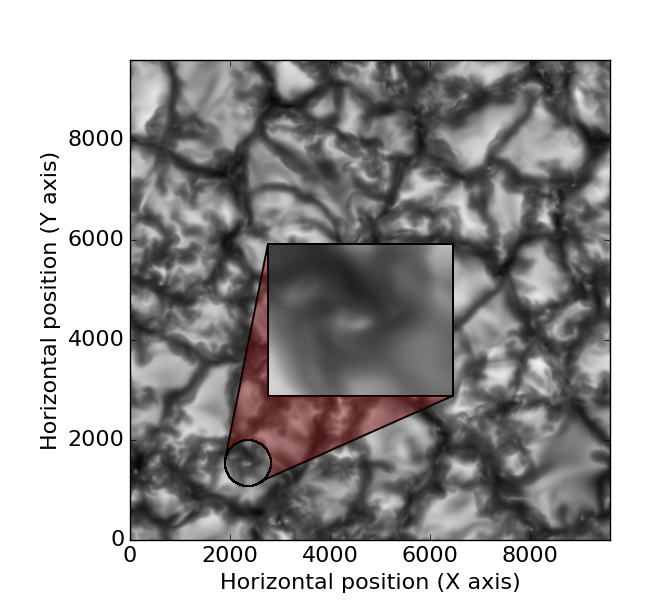}\hspace*{0.03\textwidth}
   \includegraphics[width=0.47\textwidth]{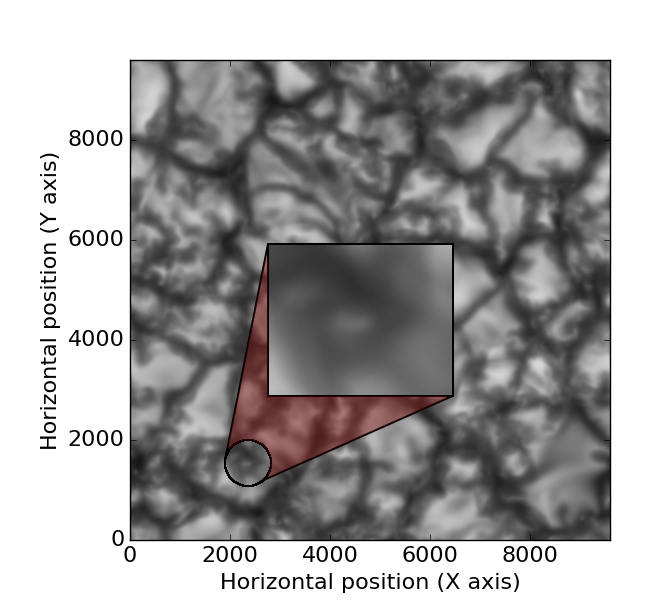}
\caption{Single snapshot of the $\mbox{Roe}_{10}$ simulation showing the intensity in the
continuum at 5000\,\AA\ (\emph{left}) and the corresponding degraded image with a GREGOR-like 
point-spread function (\emph{right}). A typical nMBP of $80\,\mbox{km}$ diameter is magnified.}
\label{fig:GREGOR_degradation}
\end{figure*}
%----------------------------------------------------------------------------

%----------------------------------------------------------------------------
\begin{figure*}
\centering
   \includegraphics[width=0.95\textwidth]{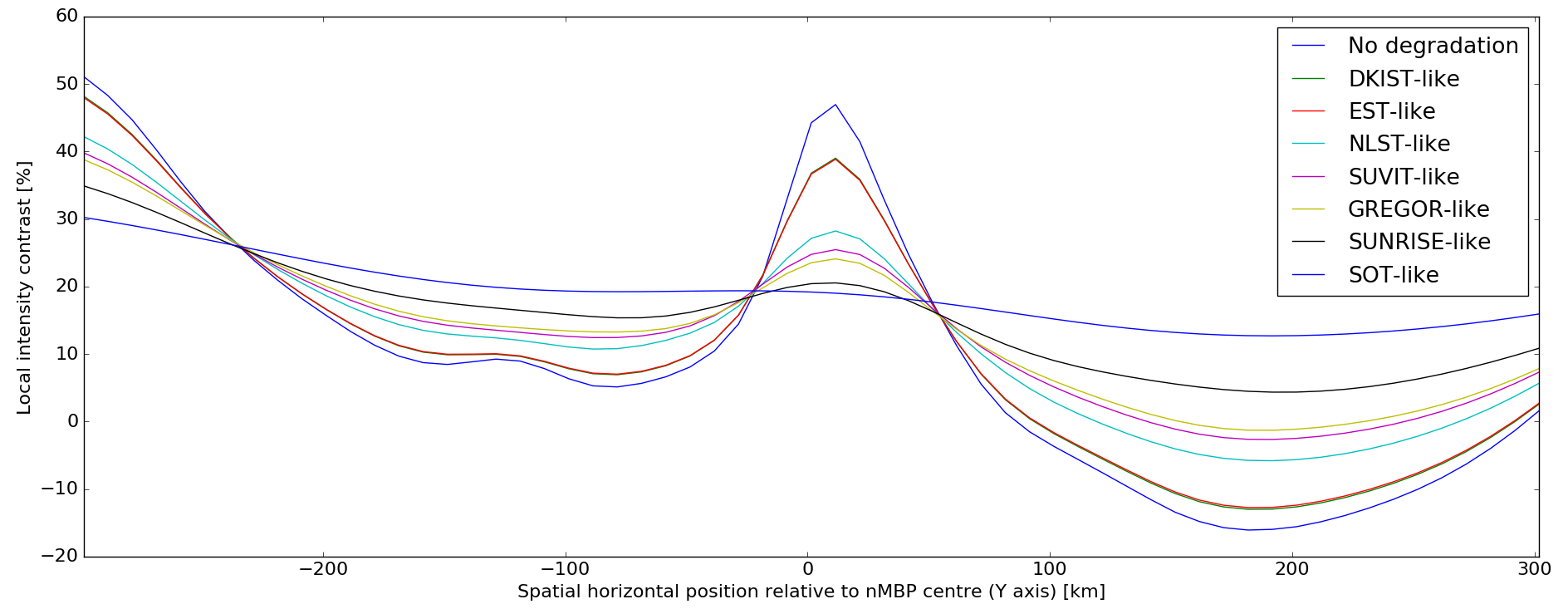}
\caption{Intensity contrast in the continuum at 5000\,\AA\  across the 
nMBP of Fig.\,\ref{fig:GREGOR_degradation}
(section along $y$-axis), using degraded images obtained with a variety of PSFs
that correspond to the solar telescopes listed in the text.
The legend orders the telescopes according to the simulated
peak contrast, from highest to lowest. The curves with degradations
corresponding to DKIST and EST are close to congruent. The coordinate
$x=0$ corresponds to the centre of the nMBP, which is taken to be the 
point of lowest density in the $\tau$\,=\,1 isosurface. Note that the peak
intensity is displaced with respect to this nMBP centre because
the nMBP funnel is inclined with respect to the vertical direction,
as is often the case, such as that visible in the right-hand panel of 
Fig.\,\ref{fig:Vertical-slices-through}.
}
\label{fig:degradations}
\end{figure*}
%----------------------------------------------------------------------------

The present bright points are non-magnetic by construction, and it is therefore unclear to what extent such bright points really exist in the solar photosphere 
that is virtually ubiquitously occupied by magnetic fields \citep{lites+al2008}. 
In reality, the pressure gradient that is at the origin of nMBPs is probably 
always provided by a combination of the centripetal force and the magnetic 
pressure gradient, however, in regions of low magnetic flux, it may happen that 
the pressure gradient results mainly from the swirling motion.

In order to gain insight into the requirements for the observational detection
of nMBPs, we have degraded our synthetic intensity maps to simulate observations with
existing solar telescopes, such as the 50\,cm Solar Optical Telescope (SOT)
aboard the Hinode space observatory \citep{tsuneta+al2008}, 
the 1\,m telescope aboard the Sunrise balloon-borne solar observatory
\citep{barthol+al2011}, and the 1.5\,m ground-based GREGOR solar telescope
\citep{schmidt+al2012}, as well as with future solar telescopes currently under construction,
such as the 4\,m \emph{Daniel K.~Inouye} Solar Telescope \citep[DKIST;][]{keil+al2010}, or planned solar telescopes such as the
space-borne 1.5\,m Solar UV-Visible-IR Telescope \citep[SUVIT;][]{suematsu+al2014},
the 2\,m Indian National Large Solar Telescope \citep[NLST;][]{hasan+al2010}, and
the 4\,m European Solar Telescope \citep[EST;][]{collados+al2010}.
The point spread functions (PSFs) that we have constructed
are convolutions of the diffraction-limited PSFs, taking the entrance
pupil (including secondary mirror and spider)
of the various instruments into account, with Lorentz functions.
The Lorentz functions are intended to account for non-ideal contributions
due to stray-light inside the instrument. Their $\gamma$ parameter
has been chosen to inversely scale with the telescope
aperture, with the reference value for the 50\,cm SOT aperture given 
by the greatest value deduced by \cite{wedemeyer2008} from
observations of the Mercury transit from 2006 and the solar 
eclipse from 2007. \cite{wedemeyer2008} has carried out convolutions 
with both Voigt and Lorentz functions in different 
observational situations. Unfortunately, the optimal parameters are 
quite sensitive to observational scenarios.
We have thus decided to retain as few parameters as possible (choosing
the Lorentz function over the Voigt function) and we also chose $\gamma=9$,
when different situations suggest a range of values from $7\pm1$
up to $9\pm1$. Our PSFs are therefore more realistic than simple
diffraction-limited PSFs but are nonetheless not fully accurate.

According to Fig.\,\ref{fig:GREGOR_degradation}, nMBPs could be seen
using the GREGOR solar telescope. However, the spatial
resolution capability of GREGOR of $0\farcs08$ at 
$500\,\mbox{nm}$ corresponds to $\approx\!60\,\mbox{km}$ on the Sun, 
so that an $80\,\mbox{km}$ nMBP would appear as a brightness 
enhancement in one single resolution element only. Furthermore, to be
sure that the nMBP harbours no strong magnetic field, a simultaneous
polarimetric measurement needs to be carried out, for which the resolution
is a mere $0\farcs3.$

From Fig.\,\ref{fig:degradations} one can see that the peak intensity 
observed in telescopes with apertures up to 2 m is barely higher 
than the intensity in the immediate vicinity, and the nMBP appears 
larger than in the raw simulations, making it very difficult to distinguish 
it from other bright structures. 

From this, we conclude that telescopes with large apertures, such as DKIST 
or EST, are needed to approach the contrast of nMBPs found in the simulations
and in order to achieve sufficiently high spatial resolution in polarimetry
for an unambiguous detection of nMBPs.

%
%%%%%%%%%%%%%%%%%%%%%%%%%%%%%%%%%%%%%%%%%%%%%%%%%%%%%%%%%%%%%%%%%%%%%%%%%%%%%
%
\section{Conclusion}
\label{sect:conclusion}
This paper investigates bright points that appear within intergranular 
lanes of high-resolution, magnetic field-free, numerical simulations of 
the solar photosphere.  
The most striking properties of these features are that on the $\tau$\,=\,1 
isosurface, their temperature is, on  average, 5\% higher than the mean 
temperature (which makes them appear bright), their mass density is lower in a funnel 
extending from the upper convection zone
to the lower photosphere, and they comprise transonic swirling motions.
At the location of the bright point, the $\tau$\,=\,1 isosurface
lies, on average, $80-100\,\mbox{km}$ deeper than the 
horizontal plane $z=0$ (corresponding to the mean optical depth
$\left\langle \tau\right\rangle =1$). At this level ($z=0$), their density
and pressure are reduced by approximately 60\% of the corresponding global average 
values at the same level. On average, their size (equivalent diameter) is approximately
$60-80\,\mbox{km}$, corresponding to 0\farcs08 -- 0\farcs11, and
their bolometric intensity contrast is approximately 20\% with respect to their
immediate intergranular surroundings. The number of nMBPs per unit area is 
0.07--0.19\,Mm$^{-2}$ and their lifespan ranges from approximately 30\,s up to the granular 
lifetime of several minutes.

Based on some of these properties an algorithm for the automatic detection 
of nMBPs has
been developed that enables us to derive statistics of their physical 
properties. Comparing the statistics of two simulations with differing 
spatial resolutions of $10\,\mbox{km}$ and $7\,\mbox{km}$, we find
nMBP equivalent diameters down to the resolution limit of the simulations.
At the upper end of the size distribution, we find twice as many bright 
points with the higher spatial resolution as we do with the lower spatial resolution.
Also, we find that in subsurface layers, swirling low density funnels 
are much more abundant than nMBPs.  These low density funnels appear only 
as bright points under the condition 
that they extend into the photosphere. The characteristics
of the nMBPs found here are in close agreement with corresponding
properties of vertical vortices found by \citet{moll+al2011}. Thus, nMBPs
are an observable manifestation of vertically directed vortex tubes, similar
to bright granular lanes, which are a manifestation of horizontally 
directed vortex tubes \citep{steiner+al2010}.

Both nMBPs and granular lanes together offer a glimpse of the elements of
turbulence at work in a stratified medium, which is of basic physical 
interest.
The nMBPs are so minuscule that they seem not to have any impact on the
overall appearance of granules and the near surface convection. However, 
as soon as magnetic flux concentrations are attracted by and caught into
the swirling down draft of a nMBP, we expect them to have a major impact 
on the tenuous atmosphere higher up through magnetic coupling 
\citep{shelyag+al2011,steiner+rezaei2012}. Chromospheric swirls
\citep{wedemeyer+rouppe2009} and magnetic tornadoes 
\citep{wedemeyer+al2012,wedemeyer+steiner2014} would be the consequences.

Such nMBPs are barely detectable using currently operating solar 
telescopes because of their small size and  relatively low contrast,
and  due to the limited spatial resolution 
of the magnetograms achievable with these telescopes (currently $\approx\!\!\!0\farcs3)$.
High-resolution 
magnetograms are needed to prove the absence of strong magnetic fields 
in nMBPs. If nMBPs exist, they should be observable with the 
new generation of 4-m-class telescopes  in regions of 
very weak magnetic fields. The Visible Tunable Filter \citep[VTF;][]{schmidt+al2014} of DKIST is designed to produce diffraction-limited magnetograms, which should therefore be adequate for 
the unambiguous detection of nMBPs.

\begin{acknowledgements}
This work was supported by the Swiss National Science Foundation under
grant ID~200020\_157103/1 and by a grant from the 
Swiss National Supercomputing Centre (CSCS) in Lugano under project 
ID~s560. The numerical simulations were carried out at CSCS on the 
machines named Rothorn and Piz Dora.
Model d3t57g44b0 was computed at the \emph{P\^ole Scientifique de 
Mod\'elisation Num\'erique}  (PSMN) at the \emph{\'Ecole Normale 
Sup\'erieure} (ENS) in Lyon.
Special thanks are extended to S.~Wedemeyer for help with the PSF 
and to the anonymous referee for very helpful comments.
\end{acknowledgements}

\bibliographystyle{aa}
\bibliography{aa28649-16}
\balance

\end{document}